%% file: cesi2018-head-research_protocol-kasia.tex
\documentclass[sigconf,anonymous=False]{cesi2018-acmart-preprint}
\usepackage{booktabs}
\usepackage[catalan,USenglish]{babel}
\usepackage{multirow}
\usepackage{caption}

\setcopyright{preprint}

\acmDOI{10.1145/3193965.3193970}

\acmISBN{978-1-4503-5736-4/18/05}

\acmConference[CESI'18]{CESI'18:IEEE/ACM 6th International Workshop on Conducting Empirical Studies in Industry}{May 28, 2018}{Gothenburg, Sweden}
\acmYear{2018}
\copyrightyear{2018}

\acmPrice{15.00}

\acmBooktitle{CESI'18:IEEE/ACM 6th International Workshop on Conducting Empirical Studies in Industry, May 28, 2018, Gothenburg, Sweden}

\begin{document}
\title[Protocol and Tools for Conducting Agile Empirical Research]{Protocol and Tools for Conducting Agile Software Engineering Research in an Industrial-Academic Setting: A Preliminary Study}

\author{Katarzyna Biesialska}
\authornote{Corresponding author is a PhD student at the Universitat Polit\`ecnica de Catalunya}
\orcid{0000-0002-2865-7990}
\affiliation{%
  \institution{CA Technologies}
  \streetaddress{WTC Almeda Park, Plaça de la Pau s/n, Edificio 2, Planta 4, Cornellá de Llobregat}
  \city{Barcelona} 
  \country{Spain}
  \postcode{08940}
}
\email{katarzyna.biesialska@ca.com}

\author{Xavier Franch}
\orcid{0000-0001-9733-8830}
\affiliation{%
  \institution{Universitat Polit\`ecnica de Catalunya}
  \streetaddress{c/Jordi Girona 1-3}
  \city{Barcelona} 
  \country{Spain} 
  \postcode{08034}
}
\email{franch@essi.upc.edu}

\author{Victor Munt\'es-Mulero}
%
\affiliation{%
  \institution{CA Technologies}
  \streetaddress{WTC Almeda Park, Plaça de la Pau s/n, Edificio 2, Planta 4, Cornellá de Llobregat}
  \city{Barcelona} 
  \country{Spain}
  \postcode{08940}
}
\email{victor.muntes@ca.com}

\renewcommand{\shortauthors}{K. Biesialska et al.}

\begin{abstract}
Conducting empirical research in software engineering industry is a process, and as such, it should be generalizable. 
The aim of this paper is to discuss how academic researchers may address some of the challenges they encounter during conducting empirical research in the software industry by means of a systematic and structured approach. The protocol developed in this paper
should serve as a practical guide for researchers and help them with conducting empirical research in this complex environment.
\end{abstract}

%
%




\newcommand{\TBD}[1]{\textbf{\sffamily\boldmath{[#1]}}}

\maketitle

\input{cesi2018-body-research_protocol-kasia}

\bibliographystyle{ACM-Reference-Format}
\bibliography{cesi2018-bibliography-research_protocol-kasia.bib} 

\end{document}

%% file: cesi2018-body-research_protocol-kasia.tex
\selectlanguage{USenglish}

\section{Introduction}
Software companies operate in a rapidly-changing environment. In such conditions traditional project management models are no longer effective, the same applies to software research management \cite{Jarvinen2017RSR}.    
Software engineering research methods need to adjust to modern times in order to accommodate fast innovation cycles and ensure that research remains relevant to industry.
Consequently, changing focus from plan-centric to feedback-centric is essential, as researchers should maximize their efforts on industry-relevant experimentation and the industrial partner's (customer) engagement. However, so far, the software engineering field has been lacking research methods and techniques, which would help with preparing a research design and conducting the actual research in industry \cite{RodriguezKO14}. In the same vein, Petersen et al. \cite{Petersen2014ARM} contend that the research community still needs to better describe its approach when conducting industry-driven research -- i.e.\ what steps it executes, the number and type of iterations it performs, among others. Furthermore, as highlighted by \cite{Petersen2014ARM}, research always depends strongly on the context, and in industry-driven research the company context is one of the most important ones to consider. However, factors such as company's culture or its organizational structure, to name just a few, usually receive little attention from software engineering researchers. And, hence, are not properly articulated in the research design and later on examined in empirical research.

Therefore, by linking action research, design science and organization theory, we address challenges that academic researchers face while conducting empirical research in the software industry these days. In this paper, we adopt the aforementioned methodologies, discuss problems with applying them and provide suggestions for improvement in the form of a protocol. Furthermore, we suggest what methods and tools
for analysis and process support can be employed by researchers in order to conduct an empirical study in an industry-academia collaborative environment in a methodological manner.
Therefore, this study can be used for diagnostic and prescriptive purposes, providing academia researchers and industry practitioners with a list of activities that should be undertaken in the course of any industrial research project being done in collaboration between a private company and a university.

This paper reports our experience in an ongoing industrial research project, which is a collaboration between academia and a global software company within the framework of an Industrial Doctorate program. Throughout the years, both partners have striven to create favorable conditions for their close collaboration, e.g.\ by being co-located at the university's campus. In the case of this particular project, the main objective is to improve Agile software development processes through a data-driven approach. The project has a duration of three years. Hence, the three major outcomes expected from it are as follows: (i) a PhD thesis completed by a student conducting the research, (ii) a quality empirical study with a positive impact on the software engineering discipline, (iii) software engineering artifacts bringing value to the company -- i.e.\ methodological guidelines and recommendations, a piece of software using the proposed methods, among others. Not only in this case, but in general, the list of potential (mostly intangible and long-term) benefits for academia and industry is longer, and may include strengthening ties between the academic community and industry practitioners, resulting in long-term collaboration prospects. In this particular example, close collaboration between partners, reinforced by co-location, allows the company not only to learn from the experts in academia, but also to frequently share and create ideas together with them. And, as a result, the company is able to position itself as a thought leader in the software engineering domain, while the university can better understand the software industry's needs and align its research interests accordingly. In general, it is said that there is a positive correlation between the presence of industry-academia collaboration and business success \cite{Lambert03}. Nonetheless, there is also a significant number of potential challenges that such partnerships can face. Garousi et al. \cite{GAROUSI2016106} have recently conducted a systematic literature study on challenges and best practices in industry-academia collaboration; therefore, it is beyond the scope of this paper to discuss in detail challenges and benefits of such a cooperation.

The remainder of this paper is structured as follows: Section 2 is meant as an introduction to the topic of action research, design science and industry-academia collaboration models. In Section 3, we discuss our protocol and describe what methods and tools software engineering researchers should embrace and exploit in order to conduct their industry-driven research in a methodological way. Finally, Section 4 serves as a summary of the paper.

\section{Background and Related Work}

\subsection{Action Research, Design Science, and an Agile Approach -- An Overview}
Many software companies have moved from plan-driven software development processes to Agile software development in the recent years. Thanks to this shift in paradigms, industry practitioners are now more likely to understand research activities driven by design science and action research, as these two approaches resemble their daily work to a large degree \cite{RodriguezKO14,Petersen2014ARM}.
Design science has been discussed during CESI (Conducting Empirical Studies in Industry) workshops several times in the recent years (e.g.\ \cite{RodriguezKO14,Wohlin201310Factors}). Action research, however, is significantly less popular in software engineering research (despite its wide acceptance in the organizational development field). Nevertheless, it is no less important from the perspective of conducting empirical research in the software industry, because action research is a research methodology used to solve practical problems by researchers and practitioners alike (usually working together as a team) \cite{Jarvinen2007Action,Petersen2014ARM}. Although both of the aforementioned paradigms have their inherent strengths and weaknesses (cf.\ \cite{Baskerville1996Critical}), they can be complementary to one another to a great extent (cf.\ \cite{Jarvinen2007Action, Hevner2010Design}).
Susman and Evered \cite{Susman1978Assessment} developed an iterative process of action research which consists of five stages and is considered a canonical one \cite{Baskerville1998Diversity}. According to Petersen et al. \cite{Petersen2014ARM}, action research resembles software process improvement and extends it with scientific guidelines and joint collaboration between researchers and practitioners. Social and organizational aspects play a vital role in action research, and hence allow to address the complexity of academia-industry collaboration. In general, action research focuses on "\textit{understanding a situation in a practical context and aims at improving it by changing the situation}" (\cite{Petersen2014ARM}, p. 55). Therefore, especially action research, and key activities performed in-line with this framework, cannot be considered in isolation from the company context. Design science, on the other hand, is more oriented towards the creation of new innovative artifacts and scientific knowledge. Nonetheless, both design science and action research have a relatively high level of abstraction and do not provide explicit guidelines or tools for researchers to follow.
The third methodology, that we want to compare, is called the Lean Research Approach for Industry-driven Research (LRA) \cite{Jarvinen2017RSR}. It has emerged quite recently to support two Finnish industry-led software research programs spanning multiple organizations. The framework is said to be the answer for Agile-driven software research programs. For the purpose of these two programs, a research road-map (SRIA) \cite{Huomo2015N4S} has been created. SRIA defines research goals and contains a list of strategic goals as well as a number of recommendations on what research activities should be conducted and how. LRA is characterized by Continuous Planning and the Research Sprint models, where Research Sprints usually last three months and their outcomes are presented during the program's quarterly review meetings (Q-reviews) \cite{Jarvinen2017RSR}.
\subsection{The Company Context -- Diagnosing the Industrial Partner's Organization}
In his seminal paper "How Do Committees Invent?" (1968), Conway \cite{Conway1968Committees} states that "\textit{organizations which design systems ({...}) are constrained to produce designs which are copies of the communication structure of these organizations}" (\cite{Conway1968Committees}, p. 31). Consequently, any piece of software produced by a company mirrors the social structure of this organization. And more broadly its organizational structure, because an organizational structure defines and facilitates the relationships between different parts of an organization in a formal way, helping to achieve organizational goals. Furthermore, organizational structures reflect formal status and power hierarchies in organizations; however, the actual patterns of information flow and collaboration are also strongly affected by their organizational cultures (including informal structures).
Therefore, in our view, software engineering researchers, when working closely with a company, should pay more attention to analyzing its organizational structure and social infrastructure, because they are reflected in the technical architecture of the software the organization produces. And, according to Conway's Law, any deficiencies in communication among the people involved in building software products in the company, hinder the product development process and result in the flawed software architectures.

Although not very popular in software engineering academic circles, expertise in analyzing the inner workings of an organization can be very helpful in empirical research (e.g.\ for obtaining access to data, project artifacts, identifying champions). Obtaining both quantitative and qualitative data is essential when conducting a comprehensive empirical study in industry. Interviewing employees and managers is often equally important as investigating the quantitative data. Both the qualitative information and the quantitative data give a researcher context and help understand the impact of her research at different levels of the organization.
Also, understanding of the actual industrial problem from a research perspective requires more than just the problem description outlined by one of a company's employees \cite{Wohlin201310Factors}. Not to mention, stakeholders' involvement in empirical studies is key. In that respect, a researcher should be able to leverage tools similar to those that external management consultants use when working with companies (described in Section 3).

\subsection{The Life-cycle and Existing Models of Industry-Academia Collaboration}
Industry-academia collaborations go through a series of phases as they evolve. Throughout the years, several models of industry-academia collaboration and technology transfer have been proposed \cite{Gorschek2006Model,Pfleeger1999Understanding,SandbergPA2011,Rombach2007Research,RunesonM2014}. In a similar vein, some researchers tried to analyze industry-academia collaboration through the life-cycle model \cite{ANKRAH2015Universities,GAROUSI2016106}. 
Although these stage models differ in the number of phases, they do share some similarities. Gorschek et al. \cite{Gorschek2006Model} presented one of the most comprehensive models, that comprises 7 phases and builds on Pfleeger's \cite{Pfleeger1999Understanding} technology transfer model.
Nevertheless, none of the discussed models contains an internal control (process improvement) mechanism which would measure the effectiveness of the process and identify potential improvement areas in the collaboration model. 

\section{Research Design}
Capitalizing on the models enumerated in Section 2, and based on recent literature reviews (e.g.\ \cite{GAROUSI2016106,ANKRAH2015Universities}) as well as our own experience, we propose a research protocol that is grounded in design science, action research and management theory, yet is capable of being used in real-life organizational settings.
Our industrial-driven research project is focused on Agile software development processes, hence the adoption of Agile-inspired  research project methodology should come naturally.
Steps outlined in Table~\ref{tab:prot} explain our protocol.

\begin{table}[!htb]
\captionsetup{font=small}
\small
  \caption{A protocol outlining stages of empirical software engineering research in industry}
  \label{tab:prot}
  \begin{tabular}{p{3cm}l}
    \toprule
    Phase & Activities in each phase\\
    \midrule
    \multirow{8}{*}{1: Define} & define the rationale for the study \\
    & define the research questions \\ 
    & define the strategy \\
    & identify the state-of-the-art \\
    & define roles and responsibilities \\
    & define vocabulary and notation \\ 
    & define the dissemination strategy\\ 
    & define the review schedule\\\hline
    \multirow{9}{*}{2: Design} & design courses of action \\
    & determine baselines \\
    & define data collection procedures \\
    & define data analysis procedures \\
    & design a solution \\
    & develop strategies for validation \\
    & design a communication plan \\
    & communicate a solution \\
    & develop a control plan \\ \hline
    \multirow{2}{*}{3: Act/Build} & design the data extraction strategy \\ & build a prototype  \\ \hline
    \multirow{2}{*}{4: Evaluate} & academia validation \\ & business validation \\
    \hline
      \multirow{2}{*}{5: Knowledge base} 
      & define dissemination mechanisms \\
      & share lessons learned \\
  \bottomrule
\end{tabular}
\end{table}

\begin{figure}[h]
\captionsetup{font=small}
\includegraphics[width=85mm,scale=0.4]{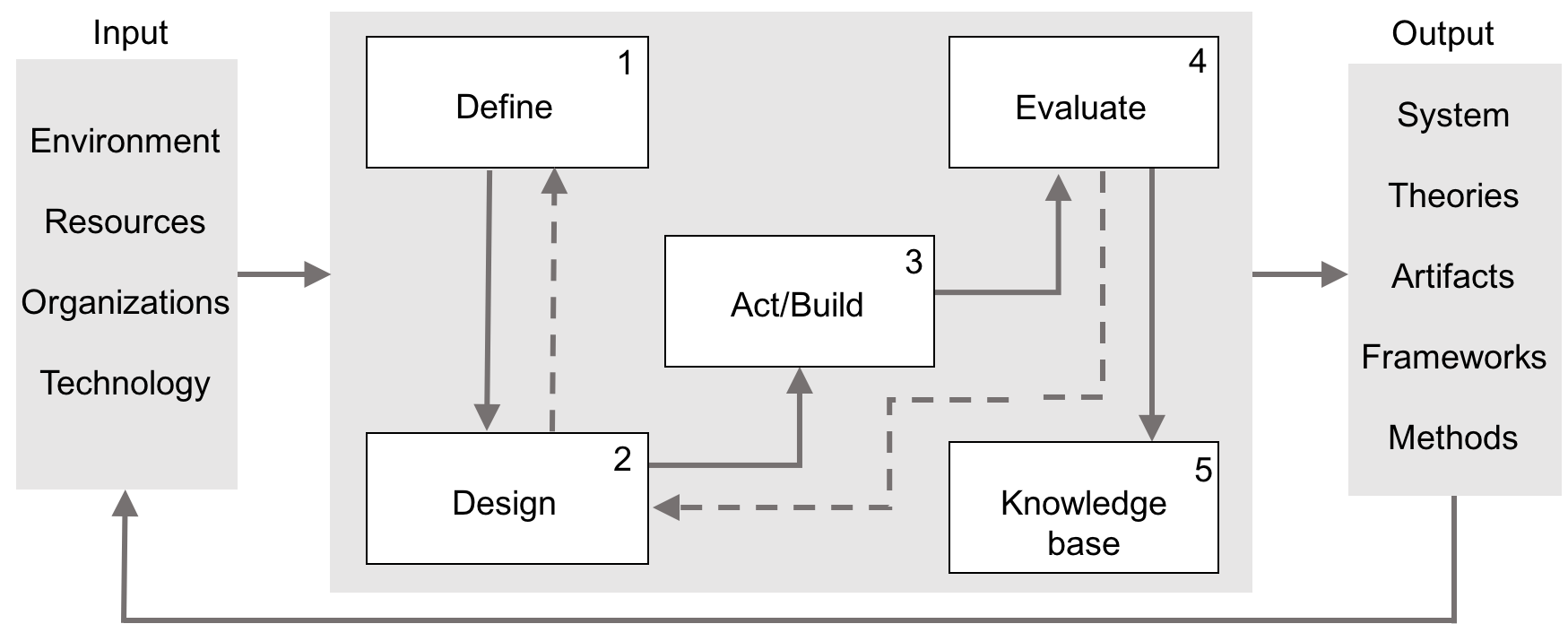}
\caption{A framework of industry-academia collaboration}
\vspace{-1.0em}
\end{figure}

Figure 1 shows our adaptation of action research and design research framework. In our case, design science is responsible for organizing research activities illustrated as an inner-loop of a software research management model (depicted as white boxes in Figure 1). Whereas action research combines all elements of industry-driven software research together and is represented as an outer-loop of industry-academia collaboration in Figure 1. Moreover, the process of industry-academia collaboration should be subject to continuous improvements, and as such can be looked through the lens of the DMAIC methodology. The DMAIC iterative cycle -- a widely used methodology in the Six Sigma world -- and its extension, the Lean SixSigma concept \cite{Staudter2009Lean6}, provide a toolbox which offers techniques that can be useful for a researcher analyzing an industrial project and improving the entire research process (mechanisms of internal control). In Table~\ref{tab:tools} we summarize activities and tools, which a researcher can exploit, to accomplish tasks from our protocol.

\begin{table} [!htb]
\small
\captionsetup{font=small}
  \caption{Tasks and tools for conducting empirical software engineering research in industry}
  \label{tab:tools}
\begin{tabular}{p{1.7cm}p{2.2cm}p{3.4cm}}
\toprule
Phase & Scope of analysis & Example of tools \\
\midrule
 \multirow{10}{1.75cm}{1: Define} & Problem & Problem frames approach \\ & Stakeholders & Stakeholder Analysis\\
& Business goals & GQM+Strategy, CTB matrix \\
& Organizational culture & OCAI, The Congruence Model, The Cultural Web \\
& State-of-the-art & SLR, mapping study, meta-analysis \\
& Responsibilities & RACI matrix \\ 
& Review schedule & Gate review \\ \hline
\multirow{4}{1.75cm}{2: Design} & Courses of action & Action Plan  \\
& Data analysis & Descriptive statistics, inferential statistics \\
& Solution & Modeling: architecture envisioning, requirements envisioning  \\ \hline
\multirow{3}{1.75cm}{3: Act/Build} & Data Extraction &  Population, sample, participants\\
& Prototype &  XP, TDD, BDD, ATDD \\ \hline
\multirow{6}{1.75cm}{4: Evaluate} & Academia validation & Peer reviews \\
& Business validation & Case studies, questionnaires, structured/unstructured interviews \\ \hline
\multirow{3}{1.75cm}{5: Knowledge base} & \multirow{3}{*}{Dissemination} & Publications, patents, conferences, trainings, workshops, seminars \\ \hline
\end{tabular}
\end{table}

Phase 1: \textit{Define}: at this stage the rationale for conducting the study from a research perspective must be defined. The research problem ought to be industry-relevant and accepted by all project partners. In our case, the rationale was prepared by the company and later on, the problem was narrowed down to address a specific business problem identified by company's customers, i.e.\ automatic dependency detection in Agile software development. Moreover, at this stage key stakeholders must be identified, so that the project gets their long-term commitment. At the beginning of the project we prepared a research plan, where we outlined its road-map, general objectives, research questions, and the dissemination strategy. The plan is frequently revised to keep all stakeholders up-to-date. In addition, short-term goals are defined in this phase. Those goals are subject to change with every iteration (as defined in the review schedule). Our project sets short-term goals on a monthly-basis. 

Phase 2: \textit{Design}: in this phase we define courses of action. Based on an initial mapping study, which was performed in the first phase, we determine an appropriate research methodology for our research and define specific research objectives. Also, a solution should be designed in collaboration with practitioners, in our case this would be an appropriate business unit. Moreover, for process improvement purposes, mechanisms for measuring the effectiveness of the whole process should be designed at this stage.

Phase 3: \textit{Act/Build}: this phase is focused on building an actual solution and executing experiments. Also, at this stage, intermediate results (such as technical reports or research papers) are published. Any solution developed in this phase is later validated and refined. For that reason, it is important to ensure that feedback cycles are fast -- e.g.\ by employing a method developed by Vetr\`o et al. \cite{VetroOF015}.  

Phase 4: \textit{Evaluate}: in this step business stakeholders provide a reality check, whereas academia validation ensures the research is methodologically sound. In order this to happen, the evaluation phase must be carefully planned in advance (Phase 2). Similar to \cite{Martinez-FernandezM14}, our protocol evaluation is based on ten factors proposed by Sandberg et al. \cite{SandbergPA2011}.

Phase 5: \textit{Knowledge base}: dissemination of research findings is done in two ways: internally and externally. Internal dissemination comprises the actual implementation of a solution in a company, knowledge sharing sessions, lessons learned. Importantly, the internal transfer of findings requires engagement from industry practitioners. Whereas external dissemination results in patents, published journal articles, and conference papers.

\section{Conclusions and Future Work}
This paper is an attempt to address the lack of Agile-driven models of the overall process of conducting empirical software engineering research in industry. The protocol developed in this work should serve as a practical guide, generalizable to most of industry-academia collaborations within the software engineering domain. We also proposed actionable points and methods that academic researchers can use when performing software engineering research in an industrial setting. Next, we plan to prepare an experience-based report evaluating how our protocol worked in practice as well as discuss lessons learned from our industrial project and industry-academia collaboration in the co-location context.

\begin{acks}
The authors would like to thank the anonymous reviewers for their valuable feedback. We also thank the Generalitat de Catalunya for support through the Industrial Doctorate grant no. 2017-DI-036. We would like to thank the European Commission for support of the Science2Society project under grant agreement no. 693651.
\end{acks}